\documentstyle[epsfig,aps,floats]{revtex}
\def\beq{\begin{equation}}
\def\eeq{\end{equation}}

\def\alphbol{\mbox{\boldmath$\alpha$}}

\def\mpi{m_\pi}
\def\mn{M_{\scriptscriptstyle N}}

\def\A{{\scriptscriptstyle A}}

\def\I{{\scriptscriptstyle I}}

\def\half{{\textstyle {1\over 2}}}

\def\smfrac#1#2{{\textstyle {#1\over #2}}}

\def\dslash{\partial\llap/}
\def\Dslash{ D \kern -0.25 cm /}
\def\ga{g_\A}
\def\ud{u^\dagger} 
\def\muicr{\mu_{\rm cr}}
\def\CPT{$\chi$PT}
\begin{document}

\title{Phases of QCD with nonvanishing isopin density}
\author{Michael C.~Birse${}^a$, Thomas D.~Cohen${}^b$ and Judith A. McGovern${}^a$}
\address{a. Department of Physics and Astronomy, University of Manchester, Manchester 
M13 9PL, UK \\
b.  Department of Physics, University of Maryland, College Park, 
20742-4111}

\maketitle

\begin{abstract}

The propagation of low-momentum baryons through QCD matter
at low isospin density is studied using chiral perturbation 
theory. When the isospin chemical potential exceeds a critical 
value, the dispersion relation for the lowest nucleon
mode becomes anomalous near zero momentum---increasing momentum 
yields decreasing energy---so that the momentum for 
the lowest energy state, $p_{\rm min}$, becomes nonzero.
This can be interpreted as a new phase of QCD, with $p_{\rm min}$ 
serving as the order parameter and the kinetic mass (defined
as the inverse of the second derivative of the energy with 
respect to the momentum at $p_{\rm min}$) serving as the 
susceptibility. A lowest-order chiral-perturbation-theory 
calculation yields a critical isospin chemical potential of 
$\sim 285$~MeV. Since this is small compared to the 
chiral-symmetry-breaking scale, corrections to it are likely
to be modest.

\end{abstract}

Recently, Son and Stephanov have considered QCD in a phase with nonzero isospin
density \cite{iso}.  Such a system is of interest for a number of reasons.  It is
an intrinsically nonperturbative system which is sensitive to much of the complex
dynamics of QCD.  In the limits of both very low and very high isospin density, 
there are analytical techniques which, given our
present state of understanding of the theory, should be both valid and computationally
tractable.  At the same time the functional determinant for the system in
Euclidean space is manifestly real and positive \cite{iso2}.  Accordingly, one can
numerically simulate QCD at nonvanishing isospin density, $\rho_\I$, (or
more precisely nonvanishing isospin chemical potential, $\mu_\I$) using standard
lattice Monte Carlo techniques, unlike the case of QCD at finite baryon
density\cite{lat}.  Thus, even if it is hard to envision a situation in which one
could study this regime experimentally, the prospect of
lattice results means that it may serve as a useful testing ground for
various theoretical approaches to QCD. The system may
also provide insights into some of the properties of neutron matter---a system
with both non-zero isospin density and baryon density.  Moreover, the phase
structure of the theory is interesting in its own right.  At low, but non-zero,
isospin density the medium is a superfluid---a pion condensate.  The phase
structure is, in fact, more rich.  For example, if one generalizes isospin to
three flavors as was recently considered by Kogut and Toublan\cite{3f} one finds a
kaon condensed phase which is separated from the pion condensed phase by a first
order phase transition.    In this paper we show that even if one restricts one's
attention to two flavors and relatively low isospin densities, the phase structure of 
the theory is quite nontrivial---in addition to the ordinary and pion-condensed phases
discussed in ref.~\cite{iso}, a new phase with radically different nucleon
properties emerges at a critical isospin density.  

Here we will focus on the low $\rho_\I$ regime.  The theoretical tool used by Son and
Stephanov to study this regime is quite conventional---chiral perturbation theory
(\CPT)\cite{chi}. The essential point is that an isospin chemical potential
couples to the quarks in QCD like a time-like photon.  Thus, by simply 
imposing gauge invariance at the level of the effective theory one has included the leading
effects of the chemical potential. Clearly, the isospin chemical potential will
have to be restricted to values that are small compared to the characteristic
hadronic scale for \CPT\ to be valid.  As noted in ref.~\cite{iso}, if the
isospin chemical potential is larger than $\mpi$ one is in a chirally nontrivial
phase.  The essential physical point is quite simple.  For $\mu_\I>\mpi$ it
becomes energetically favorable to condense pions out of the vacuum.  Pions
continue to condense until the pion-pion interactions (which can be described
accurately by \CPT) yields a repulsion equal to the attraction due to the
chemical potential.  Thus as the isospin chemical potential is increased from
zero, the isospin density remains zero until $\mu_\I=\mpi$ at which point $\rho_\I$
increases smoothly from zero.  From the perspective of  \CPT, which is based
on a nonlinear sigma model, the pion condensation described as a chiral
rotation from the isoscalar ($\sigma$) direction toward one of the pion directions.  

At first sight it might appear that this pion-condensation transition is the only
one which can be described in a regime where \CPT\ is valid.  After all,
it is necessary to have an order parameter which
distinguishes the pion condensate from any putative new phase.  In the
context of the chiral Lagrangian---a simple theory of interacting pions---there is
no obvious candidate for such an order parameter aside from the chiral field
itself.  However, pions are not the only possible
hadronic probes of this state of matter:  one can send nucleons through this
matter to probe its structure.  

The nucleon dispersion relation can be used to define a new phase.  For nucleons
in free space, Lorentz invariance completely determines the dispersion relation:
$E^2 = p^2 + \mn^2$.  However, an isospin chemical potential breaks
Lorentz invariance since it couples to the zeroth component of
a four-vector.  Thus, the nucleon dispersion relation in this medium can be
nontrivial.  At sufficiently low $\mu_\I$, both above and below the pion
condensation transition, the nucleons have a normal dispersion relation---the
energy increases with increasing momentum.  However, when $|\mu_\I|$ is above a
certain critical value, $\muicr$, the situation is fundamentally
different; for low values of the nucleon momenta the dispersion relation for the
lowest lying nucleon excitation is anomalous in the sense of anomalous dispersion
in optics \cite{Jackson}---increasing the momentum leads to lower energy.  This defines our
new phase.  It can be described by a simple order parameter,
 \begin{equation} \Theta \, =
\, p^{\rm baryon}_{\rm min}\, ,
\label{order} \end{equation}
where $p^{\rm baryon}_{\rm min}$ is the magnitude of the momentum of the lowest-lying 
excitation of nonzero baryon number.  For $|\mu_\I| < \muicr$,  $\Theta=0$ while 
for $|\mu_\I| >\muicr$, $\Theta \neq 0$; thus $\Theta$ serves as an order parameter in
the traditional sense of distinguishing between two phases.  As will be shown
below, the critical chemical potential, $\muicr$, can be estimated using
\CPT.  At lowest order in the chiral expansion the critical chemical potential satisfies

 \begin{equation}
   \,\frac{2\,\mn\,\mpi^2 \, \muicr}{\left( 1 - \ga^2 
\right) \,\mpi^4 +\ga^2\, \muicr^4} \, = \, 1\, ,
\label{crit} \end{equation}
 which yields an estimate of $\muicr\sim 285$~MeV for the transition.  
Chiral corrections to this are of order $Q^2/\Lambda^2$, so that one might expect
corrections to alter this value by $\sim 20 \%$.  

We begin with a brief recapitulation of the arguments of ref.~\cite{iso} for the low
$\mu_\I$ region.  As all dynamical scales in the problem are much less than typical
hadronic scales one can describe the system by an effective theory
for interactions among pions, the only light degrees of freedom in the
problem.  The Lagrangian for the purely pionic part of the theory is simply that
of the nonlinear sigma model.  At leading order it is
\begin{equation} {\cal L} = {1\over 4}f_\pi^2 \, {\rm Tr}\left 
[\nabla_\mu U \nabla^\mu U^{\dagger} + 2\mpi^2 {\rm Re}\,U \right] \, ,\label{Lg} 
\end{equation}
where 
$U \in SU(2)$ and can be parametrized by $U \equiv \exp \left ( i \frac{\vec{\pi}
\cdot \vec{\tau}}{f_\pi} \right )$.
The theory at this order contains two
parameters: the pion mass, $\mpi \approx 139$~MeV, and the pion decay constant
$f_\pi \approx 93$~MeV; the covariant derivative $\nabla$ incorporates couplings 
to gauge fields.

At the QCD level the isospin-chemical-potential term is simply
$\half\mu_\I v_\mu \overline{q} \gamma^\mu \tau_3 q$, where $v_\mu$ is a four velocity
that describes the motion of the resulting medium; in its rest frame  
$v_\mu= (1,0,0,0)$. Thus, the isospin chemical potential couples directly to the 
conserved vector-isovector current, and at the level of the effective theory, 
the coupling of the isospin chemical potential must be the same as the coupling of 
a U(1) isovector gauge field.\cite{iso}.  Hence we use
\begin{equation}\nabla_\mu U = \partial_\mu U +i[U,V_\mu], \end{equation} 
with 
$V_\mu=\half\mu_\I v_\mu \tau_3=\half\mu_\I \tau_3 \delta_{0\mu}$. Using this lowest 
order Lagrangian one expects predictions of pionic observables and thermodynamic 
quantities to be
accurate up to corrections of relative order $Q^2/\Lambda^2$ where $Q \sim \mpi,
\mu_\I, p$ (where $p$ is the external momentum of a probe and 
$\Lambda \sim 1$~GeV is a characteristic hadronic scale).

The next step is to determine the  $U$ which minimizes the energy.  The
ground state is translationally invariant and static so one can simply use a
constant $U$ in the Lagrangian of eq.~(\ref{Lg}) and equate it to
the negative of an effective potential,
\begin{equation}
  V_{\rm eff}(U) 
= {f_\pi^2 \, \mu_\I^2 \over8} {\rm Tr}  ( \tau_3 U \tau_3 U^{\dagger} \, - \, 1 )
  - {f_\pi^2 \, \mpi^2\over2}{\rm Re} \, {\rm Tr} \, (U) .
  \label{pot}
\end{equation}
It is clear that first term in eq.~(\ref{pot}) lowers the energy if $U$ aligns
along the $\tau_1$ or $\tau_2$ directions while the last term favors the isoscalar
direction.  Accordingly, the minima can be found using an ansatz which allows for
general rotation between the isoscalar and $\tau_1$ or $\tau_2$ directions,
\begin{equation}
 U = \cos\alpha \, + \, i (\tau_1\cos\phi + \tau_2\sin\phi)
\sin\alpha\ , \label{Ansatz}
 \end{equation} 
for which the  effective potential takes the form 
\begin{equation} 
V_{\rm eff}(\alpha) = {f_\pi^2 \, \mu_\I^2\over4}(\cos 2\alpha \, - \, 1) -
f_\pi^2 \, \mpi^2 \cos\alpha\ .  \label{Valpha}
 \end{equation} 
Note that $V_{\rm eff}$ depends on $\alpha$ but not on $\phi$.  
Minimizing with respect to $\alpha$
one sees that for $|\mu_\I| < \mpi$ ,  the minimum occurs at $\alpha=0$ ($U= 1$) so
that one is in the normal phase.  For $|\mu_ I| > \mpi$, the minimum occurs at\cite{iso}
\begin{equation} \cos\alpha={\mpi^2\over \mu_\I^2}.  \label{alpha} \end{equation}
which implies that the state has chirally rotated to a new pion-condensed state.

Now let us turn to the main subject of the present work, the propagation of nucleons 
through a pion-condensed phase. The interaction between nucleons and the chiral field 
can also be described
using \CPT.  The lowest order Lagrangian describing the interaction of nucleons with 
pions in \CPT\ is given by\cite{chi} 
\beq
{\cal L}_{\pi N} \, = \, \Psi \left ( i{\Dslash}-M+\half \ga \gamma^\mu\gamma_5 
u_\mu \right )\Psi
\, ,
\label{Lpn}\eeq
where $u^2=U$ and
\begin{eqnarray}
D_\mu&=& \partial_\mu+\half[\ud,\partial_\mu u]-\half i (\ud V_\mu u+u V_\mu\ud)\, ,
\nonumber\\
u_\mu&=&i(\ud\nabla_\mu u -u\nabla_\mu \ud)\, ,\nonumber\\
\nabla_\mu u &=& \partial_\mu u +i[u,V_\mu]\,.
\end{eqnarray}
We note that the interaction between nucleons and pions considered by Son and 
Stephanov\cite{iso}
was based on a nucleon field that transforms linearly under 
chiral transformations rather than one that transforms nonlinearly 
as considered here. 
In fact, it is straightforward to show that the two are equivalent---that is,
they are  related to each other by a change of variables---provided that one takes 
$\ga =1$.
The form used in eq.~(\ref{Lpn}), the standard form used in baryon \CPT,
has the advantage of being able to incorporate  $\ga \neq 1$  in a straightforward 
way.  As detailed above, the isospin chemical potential acts like a vector field:
$V_\mu=\half \mu\tau_3 \delta_{\mu 0}$.  Inserting this into the Lagrangian
and varying with respect to $\overline{\Psi}$ yields the following Dirac equation:
\begin{equation}
\left [ i\dslash \,- \, M \,  +  \, \half\mu\gamma_0
\left(\half (u \tau_3 \ud +\ud \tau_3 u) 
 \, +  \, \ga\gamma_5\half (u \tau_3 \ud -\ud \tau_3 u)\right) \right ] \Psi \, = \, 0
\, .
\end{equation}
With the ansatz of eq.~(\ref{Ansatz}) for the pion field, and a plane-wave state for
the Dirac field, this gives
\beq
\left [ E-\alphbol\cdot{\bf p} \,- \, M\gamma_0 \,  + \, 
\half\mu\left(\cos\alpha \,\tau_3 \,+\,\ga \gamma_5\sin\alpha\,
(\cos\phi\,\tau_2-\sin\phi\,\tau_1)\right)\right ]\Psi \, = \, 0\, .
\eeq
The solutions are eigenstates of
helicity, and  there are four doubly-degenerate solutions for each value of the 
momentum.   The two positive-energy solutions are given by 
\beq
E_{\pm}=\sqrt{M^2 + p^2 + \smfrac 1 4 \mu_\I^2\,
        (\cos^2  \!\alpha + 
     \ga^2\sin^2 \!\alpha) \pm | \mu_\I |\,\sqrt{M^2\,
           \cos ^2 \!\alpha + 
          p^2\,(\cos^2  \!\alpha + 
          \ga^2\,\sin ^2 \!\alpha)}}\, .
\label{eps}\eeq
Although the pion condensate breaks parity, it also breaks CP and thus
conserves charge conjugation.  Thus, the energies of the positive- 
and negative-energy solutions are equal up to an overall minus sign. The pion-condensed
phase breaks $I_3$ since the condensate is aligned in one direction in the
$\tau_1-\tau_2$ plane.  Thus, in this phase the solutions of the Dirac equation
do not correspond to states of well-defined isospin---rather they are
admixtures of $I_3 = \pm 1/2$.  Only for $\alpha=0$ are the solutions eigenstates of
$I_3$.
  
\begin{figure}[t]
 \begin{center} \mbox{\epsfig{file=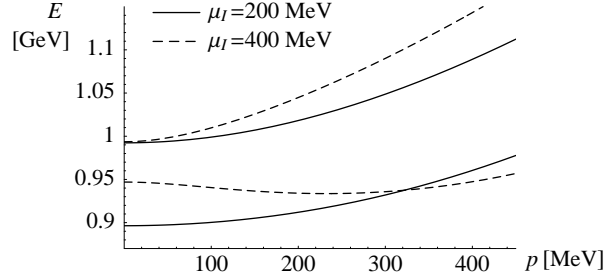,width=8truecm,angle=0}}
  \end{center}
\vskip -0.7cm
\caption{The nucleon dispersion relation for values of $\mu_\I$ below and 
above $\muicr$.
\label{dispfig}}
\end{figure}

Ref.~\cite{iso} considered  the  ``nucleon mass'', 
defined as the minimum energy needed to produce a baryon-number-one excitation, 
implicitly taken to be at $p=0$.  Their expression corresponds to the first two terms
in the expansion of eq.~(\ref{eps}) in powers of $1/M$.
However looking at the full dispersion relation, we see that for $\alpha$ close
to $\pi/2$ the point $p=0$ is no longer the minimum for the lower branch.
Expressing $\alpha$ in terms of $\mu_\I$ through eq.~(\ref{alpha}), we find a critical
value of $\mu_\I$, already given in eq.~(\ref{crit}), above which the minimum energy
is at a non-zero value of $p$. 
Fig.~\ref{dispfig} illustrates the qualitative difference between the two regions. We
plot $E_\pm$ for two choices of $\mu_\I$, one below the
critical value and one above.  

\begin{figure}[t]
\begin{center} \mbox{\epsfig{file=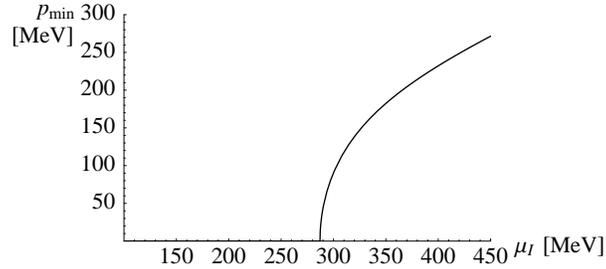,width=8truecm,angle=0}}
  \end{center}
\vskip -0.7cm
\caption{The value of $p$ which minimizes the energy as a function of $\mu_\I$, 
eq.~(\ref{pmin}).
\label{pminfig}}
\end{figure}

As noted earlier, $p_{\rm min}$ functions as an order parameter. For $|\mu_\I|>\muicr$
it is given by
\beq 
p_{\rm min} \, =  \,
\frac{1}{2} \, \sqrt{\frac
{\left(\ga^2\mu_\I^4\,+\,(1-\ga^2)\,\mpi^4\right)^2\,-\,4M^2\mpi^4\mu_\I^2}
{ \mu_\I^2 \, \left(\ga^2 \mu_\I^4 \, + \,  (1 -\ga^2 ) \, \mpi^4 \right )}}\,.
\label{pmin}
 \eeq
This is plotted in fig.~\ref{pminfig}.
  
\begin{figure}[t,h]
\begin{center} \mbox{\epsfig{file=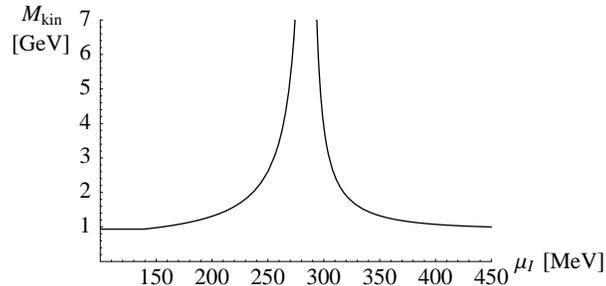,width=8truecm,angle=0}}
  \end{center}\vskip -0.7cm
\caption{The kinetic mass as a function of $\mu_\I$, eq.~(\ref{mkin}).
\label{mkinfig}}
\end{figure}

An alternative definition for the mass to that considered in ref. \cite{iso}
is the kinetic mass, which for any mode is defined by 
\beq
M_{\rm kin}^{-1} \, = \,  \left . \frac{\partial^2 E}{\partial p^2} 
\right |_{p_{\rm min} \; .}  
\eeq
Clearly for the free-space dispersion relation the mass defined by the minimum-energy 
excitation and the kinetic mass are equal, but in medium this need not be
so.  The kinetic mass is important in that it is the inertial parameter which
enters in an effective Schr\"odinger equation describing the lowest-energy excitations 
of this system.  The kinetic mass for the lower of the two
nucleon modes is 
\beq
M_{\rm kin}  =  \left\{\begin{array}{r@{\quad\quad}l}
\rule[-3mm]{0mm}{12mm}
\displaystyle \frac{M \, \mpi^2 \, \sqrt{ (1 -\ga^2) \, \mpi^4 
\, + \, \ga^2 \mu_\I^4\,- \, 4 \, M \, \mpi^2 \mu_\I \, + \, 4 M^2 \mu_\I^2 } }{ 2 M
 \mpi^2 \mu_\I \, - \,  (1- \ga^2) \, \mpi^4 \, - \, \ga^2 \mu_\I^4} 
& |\mu_I|<\muicr\, ,\\  \rule[-3mm]{0mm}{12mm}
 \displaystyle \frac{ \ga M\bigl((1-\ga^2)\,\mpi^4+\ga^2\mu_\I^4\bigr)^{\frac 3 2}
\sqrt{\mu_\I^4\, -\, \mpi^4}}
{\left(\ga^2\mu_\I^4\,+\,(1-\ga^2)\,\mpi^4\right)^2\,-\,4M^2\mpi^4\mu_\I^2}
& |\mu_I|>\muicr\, .
\end{array}\right.
\label{mkin} 
\eeq
It is plotted in fig.~\ref{mkinfig}. One striking feature is that $M_{\rm kin}$
develops a  pole, diverging at the critical value $\muicr$.   This critical value
occurs when the denominators of eq.~(\ref{mkin}) vanishes. 

Figures~\ref{pminfig} and \ref{mkinfig} look very much like a standard 
second order phase
transition, with the kinetic mass diverging at the transition in the manner of a
typical susceptibility and $p_{\min}$ rising from zero as a fractional power in
the manner of a typical order parameter.  In a way, however, this phase
transition is rather unusual.  Generally at phase transitions there are
discontinuities in quantities associated with bulk energetics such as specific
heats.  In the present case we have no evidence of such effects.  The basic
reason is very clear: the discontinuities associated with the phase transition
reported here all involve baryon properties.  To the order at which we have 
calculated, the pionic degrees of freedom are unaffected by the transition and the 
pions dominate the energetics.  To the extent that there are discontinuities in 
quantities associated with energies they must involve changes in possible 
nucleon-antinucleon loop effects which might altered discontinuously when the nucleon 
dispersion relation is altered.  However, any such possible effects are too small
to be calculated in \CPT.

We note that the value of the critical chemical potential from Eq.~2 is of order $(\mn\mpi^2)^{1/3}$, and numerically
equal to 285~MeV$\approx 2\mpi$.  Thus in practice $\muicr$ is comparable in size to $\mpi$ and well below $\mn$, 
so we are in the
usual regime of applicability of \CPT.  The leading corrections to our result should come from the inclusion of the 
higher-order terms in the pion-nucleon Lagrangian which are parameterised by the low-energy constants $c_i$.  In fact 
to order $1/\mn$ the sole effect of these is to add a term  $(4c_1 \mpi^2 - (c_2 + c_3)\mu^2 )\sin^2\alpha$ to the 
nucleon energy.  Not only is this small ($58\pm1$~MeV at $\mu=\muicr$ using the second-order determinations of the 
$c_i$ \cite{fettes}), it is also momentum-independent.  Thus it cannot affect $\muicr$ at all, and corrections
start only at third order.

Finally, we turn to a question of considerable significance in nuclear physics,
namely neutron matter, which is QCD at finite baryon chemical potential and
isospin chemical potential with the chemical potentials chosen so the that baryon
density is twice the isospin density.  It is unclear whether the \CPT\
techniques used here can be generalized easily to deal with this problem since they are
based on the assumption that the only light
scale in the problem is the pion mass.  On the other hand, the inverse of the
nucleon-nucleon scattering length, $1/a$, defines a mass scale much lighter than
$\mpi$\cite{wein}.  This suggests that one should develop a systematic expansion treating
both $1/a$ and $\mpi$ is light.  Formally, such expansions have been developed
for the two nucleon problem \cite{ksw} however, in that context there is clear
evidence for problems with convergence of the chiral expansion\cite{conv}. However,
the effects of chiral symmetry may be very different in the case of neutron matter
from those in the N-N scattering case and a viable expansion may prove possible.  The
possibility that at some density neutron matter exhibits pion condensation is
interesting since
the fact that the propagation of nucleons through a pion condensate changes
characteristically above a certain isospin chemical potential could play a
significant role.

With the above caveats, we have looked at the mean-field Fermi gas in the present
framework.  The possibility of pion condensation arises from the term 
$-\half \mu_\I^2 f_\pi^2 \sin^2\!\alpha$ in the energy, which  for sufficiently
large $\mu_\I$ can overcome the mass term $- \mpi^2 f_\pi^2 \cos\alpha$; the question
is whether a gas of fermions tends to stabilize or destabilize
the phase with condensed pions. We see
from  eq.~(\ref{eps}) that for a given value of $\mu_\I$, the lowest 
single-nucleon energy is at $\alpha=0$, the normal vacuum.  Analytic calculation
shows that for fixed $\mu_\I$ and baryon density, $\alpha=0$ is also a minimum of 
the full fermion energy, whereas $\alpha=\half \pi$ is a maximum.  Thus the
fermions do not favor the pion-condensed phase.  Numerical calculations were done
for the case with the isospin density constrained to equal minus half the baryon 
density, as in neutron matter. In the chiral limit $\alpha=\half \pi$  
(which minimizes the pion energy) is a minimum of the full pion and fermion energy only 
for baryon densities below about a third of nuclear matter density ($\rho_0$).  
For higher densities the minimum moves round the chiral circle, reaching $\alpha=0$ 
at about  two-thirds of $\rho_0$. Thus even in the chiral limit the 
condensed phase is a low-density phenomenon. With the physical pion mass, pion 
condensation requires  $\mu_\I>\mpi$ at zero baryon density; in neutron matter such
a value is reached at around $4\rho_0$. Thus the Fermi gas model does not predict 
this form of ($s$-wave) pion condensation, much less the new phase discussed in this 
paper, in neutron stars.

\medskip

TDC gratefully acknowledges the support of U.S.~Department of Energy via
grant DE-FG02-93ER-40762.  MCB and JMcG are supported by the UK EPSRC, which also
funded a visit by TDC  to Manchester during which the  current work was started.
We would like to thank D. T. Son and M. A. Stephanov for their helpful comments.

\end{document}